\documentclass{sigchi}
\usepackage{algorithmic,algorithm}
\usepackage{ulem}

\usepackage{balance}       
\usepackage{graphics}      
\usepackage[T1]{fontenc}   
\usepackage{txfonts}
\usepackage{mathptmx}
\usepackage[pdflang={en-US},pdftex]{hyperref}
\usepackage{color}
\usepackage{booktabs}
\usepackage{textcomp}

\usepackage{microtype}        
\usepackage{ccicons}          

\usepackage{todonotes}

\def\plaintitle{Evaluating Voice Skills by Design Guidelines Using an Automatic Voice Crawler}

\def\emptyauthor{}
\def\plainkeywords{intelligent voice assistants; voice user interface design; user experience evaluation;}

\makeatletter
\def\url@leostyle{%
  \@ifundefined{selectfont}{
    \def\UrlFont{\sf}
  }{
    \def\UrlFont{\small\bf\ttfamily}
  }}
\makeatother
\def\pprw{8.5in}
\def\pprh{11in}

\definecolor{linkColor}{RGB}{6,125,233}
\hypersetup{%
  pdftitle={\plaintitle},
  pdfauthor={\emptyauthor},
  pdfkeywords={\plainkeywords},
  pdfdisplaydoctitle=true, 
  bookmarksnumbered,
  pdfstartview={FitH},
  colorlinks,
  citecolor=black,
  filecolor=black,
  linkcolor=black,
  urlcolor=linkColor,
  breaklinks=true,
  hypertexnames=false
}

\begin{document}

\title{\plaintitle}

\numberofauthors{3}
\author{%
  \alignauthor{Xu Han\\
    \affaddr{Boulder, USA}\\
    \email{xu.han-1@colorado.edu}}\\
  \alignauthor{Tom Yeh\\
    \affaddr{Boulder, USA}\\
    \email{tom.yeh@colorado.edu}}\\
}

\maketitle

\begin{abstract}
Currently, adaptive voice applications supported by voice assistants (VA) are very popular (i.e., Alexa skills and Google Home Actions).
Under this circumstance, how to design and evaluate these voice interactions well is very important. In our study, we developed a voice crawler to collect responses from 100 most popular Alexa skills under 10 different categories and evaluated these responses to find out how they comply with 8 selected design guidelines published by Amazon. Our findings show that basic commands support are the most followed ones while those related to personalized interaction are relatively less. There also exists variation in design guidelines compliance across different skill categories.
Based on our findings and real skill examples, we offer suggestions for new guidelines to complement the existing ones and propose agendas for future HCI research to improve voice applications' user experiences.
\end{abstract}



\section{Introduction}
Voice applications running on voice user interface (VUI) devices have recently achieved significant commercial success.
In the USA, 47.3 million (19.7\% of) households now own VUI devices (2018), that is up from less than 1\% of the population just two years ago \cite{Voice18}.
71.9\% of those devices are Amazon's Echo devices, followed by Google's devices with 18.4\% \cite{Voice18}.
China is another country with a rapidly growing market projected to reach \$23 billion this year, according to Juniper Research\cite{Jresearch}.

One key characteristic of this new generation of VUI devices is that they provide an API platform for third-party developers to design and build voice applications and publish them on a marketplace with the potential to reach millions of users.
Amazon's Alexa skills \cite{Alexaskill} and Google's Google Home Actions \cite{Googleactions} are the two most popular examples. 
However, many third-party developers may not have prior experiences in designing and building voice applications, especially in terms of user experience (UX). 
To help educate those developers, Amazon and Google have published design guidelines \cite{Alexa2018-rd, GuideG} to establish a set of good UX design practices a voice application should try to comply with.
These official design guidelines cover a variety of UX topics ranging from how to clearly communicate the purpose of a voice application to users to how to design a natural and adaptive interaction flow.

There is a huge body of literature in HCI that propose design guidelines to educate practitioners in the field who want to design and develop an application for a wide range of interactive technologies (i.e. web readability design \cite{Miniukovich2017-kd}, gesture user interface design \cite{Gheran2018-od}).
However, most of these research efforts were concluded at the publication of these guidelines; few went further to understand whether these guidelines would be later on accepted and followed by designers and developers in the wild.
In the example of Amazon, design guidelines for Echo, crafted by its own team of UX researchers, have been published for more than a couple of years \cite{Alexa2018-rd}. 
Tens of thousands of developers have designed and published voice skills by following them.
This situation provides a good opportunity to study the adoption pattern of design guidelines in the wild.

An example of a well-designed Alexa skill is \textit{Would You Rather for Family}. This skill is an interactive Q\&A game that exhibits several design features following Amazon's guidelines, 
including remembering where the last interaction ends, giving a personalized opening prompt to users, and speaking naturally. 
Deservedly, this skill has a high average rating -- 4.9 out of 5 stars based on 3209 user reviews.
In contrast, an example of a poorly-designed Alexa skill is \textit{AccuWeather}.
This skill's average rating is low -- 2.2 out of 5 stars based on 182 user reviews.
By interacting with this skill, we can tell that the skill's design violates several design guidelines, such as handling errors properly. 
These violations are also complained by some users in their reviews.
By analyzing a large number of skills like these, we can gain insights into design guidelines' adoption pattern.
We want to ask: \textbf{Among the design guidelines for voice applications, which are followed or violated more often by developers in the wild? (RQ1)}
Another phenomenon we observed is the high variance in user ratings across app categories. 
For instance, we found the average user rating of top 10 popular skills in the Games category is 4.5, comparing to 2.6 for those in the Food \& Drink category. Motivated by this phenomenon, a second research question can be opened: 
\textbf{Could this high degree of variability among categories be related to whether certain guidelines are followed or not followed? (RQ2)}  


To study these questions, we decided to limit the scope to  Alexa skills in this paper.
We selected a sample of 100 most popular Alexa skills from ten different categories and evaluated whether their designs follow the a selected subset of Amazon's official design guidelines.
Note that our scoping decision does not imply an acknowledgement of Amazon's design guidelines as the gold standard nor an endorsement of Amazon's products. Rather, the decision is based on where we might be able to gather the most data, which platform has the largest number of developers in the wild, and which set of design guidelines are most likely read by these developers (which is unlikely an academic paper).
To automate our data collection process, 
we developed a voice skill crawler to collect responses from these skills under different commands input.
We then analyzed the collected responses to determine whether or not certain guidelines are followed.
Regarding the first research question, an example of key findings is that basic commands support are the most obeyed guidelines while personalized service-related guidelines are relatively less obeyed.
Regarding the second research question, an example of key findings is that skills in the Games category on average obey the most guidelines while skills in the Entertainment category the least.

\noindent Furthermore, previous research (e.g. \cite{Luger2016-pd, Cowan2017-zc}) has studied the general gulf between user expectations and real user experiences, which indicates a need of a comprehensive set of UX design guidelines for developers. Whilst UX design guidelines exist (e.g. Amazon and Google design guides), further revision iterations are still necessary. 
Thus, based on the findings on a large sample of skills in our evaluation process, we identified several aspects that current UX design guidelines do not cover and proposed additional design recommendations to fill this gap.
In the remainder of the paper, we provide related work, a detailed description of our method, a comprehensive presentation of our findings regarding the two research questions, suggestions for how to improve the current design guidelines, and agendas for future HCI research.


\section{Related Work}

\subsection{Limitations on User Experience of VUIs}
Recent years' advances in speech technology have led to voice user interfaces' (VUIs) improved accessibility and they have been studied in the HCI literature in a wide variety of application contexts (e.g. assistive services \cite{Chen2015-lb}, education \cite{Smith2016-zu}, health \cite{Damacharla2018-in,Brown2017-ao}, entertainment \cite{McReynolds2017-zx, Purington2017-ky} and Internet of Things (IoT) \cite{Fytrakis2015-td,Braun2017-yo}). However, despite the benefits and convenience they have brought with us, VUIs still possess several limitations that would affect the user experience (UX).     
Some users may feel less in control since VUI provides no visual feedback \cite{Luria2017-uz} and the lack of VUI system transparency would result in users either feeling overwhelmed by the unknown potential, or led them to assume that the tasks they could accomplish were highly limited \cite{Luger2016-pd}. 
In some situations, voice interactions may evoke negative feelings in users \cite{Luria2017-uz}. 
In terms of subjective satisfaction, users may not feel comfortable talking with machines if the synthesized speech does not sound natural \cite{Zou2015-jn}. 
More specific to voice assistants (VAs) like Amazon's Alexa, several issues have been reported,
such as concerns over users' privacy \cite{Saffarizadeh2017-tm,McReynolds2017-zx}, technical limitations of natural language processing \cite{Zou2015-jn} and restricted communication protocols \cite{Vtyurina18}. Under these circumstances, user experience evaluations of VUI, specifically VAs, deserve further attention and studies.

\subsection{User Experience Evaluation of VUIs}

Based on the results of our literature survey, we noted several existing user experience evaluation methodologies that could be applied to VUIs or VAs. 
Traditional usability studies are very useful in gathering feedback and conducting evaluation analysis. In \cite{Luger2016-pd}, researchers interviewed 14 users of VAs in an effort to understand the factors affecting everyday use. 
\cite{Ali-Hasan18} deployed traditional lab-based usability studies using multiple fidelities like static mock, functional prototype and launched products for future design iterations. 
At the same time, longitudinal study is another effective methodology that can shed lights on real life scenarios and situations for using VAs\cite{Ali-Hasan18}.

Specifically for Alexa skills' user experience evaluation, although Alexa provides an overall platform for developers to check their skills before submitting to the review process, there is still no guarantee these skills follow the published voice design guide \cite{AA18}.
The user experience evaluation of skills still heavily relies on subjective data such as user ratings, reviews, feedback and reports. 
Thus, there is a need for a more systematic and objective approach to evaluating voice skills.
Our study represents one possible approach by comparing across a large number of voice skills and examining their designs with respect to official design guidelines.

\section{Method}
In order to investigate the adoption and compliance pattern of current Alexa design guidelines, we first developed a crawler system to collect responses from a sample of 100 Alexa skills and then manually labeled those collected responses to study whether or not they comply with the selected design guidelines.
Here we elaborate our method in details.

\subsection{Alexa Skills Selection}
More than 30,000 Alexa voice skills \cite{Voiceskillnumber} have been published by thousands of third-party developers.
On Alexa's website, these skills are organized by categories. 
Because the variance on user average ratings across different categories is high, we are interested in studying these skills.
In this case, we wanted to collect a representative sample for the purpose of our research.
First, we identified the ten top categories with the most number of skills.
The ten categories (and their subcategories) are: 1. Daily Activities (News, Weather), 2. Entertainment (Movies \& TV, Music \& Audio, Novelty \& Humor, Sports), 3. Education \& Reference, 4.Health \& Fitness, 5.Travel \& Transportation, 6.Games, Trivia \& Accessories, 7. Food \& Drink, 8. Shopping and Finance (Shopping, Business \& Finance), 9. Communication and Social and 10. Kids. 
We wrote a script to scrape Alexa's website to pick the top ten skills for each category based on the number of reviews. 
For categories with subcategories, we tried to balance the number across the subcategories manually.
For example, the ten skills we selected to represent the Entertainment category consist of
three in the Movies \& TV subcategory, three in the Music \& Audio subcategory, two in the Novelty \& Humor subcategory, and two in the Sports category. 
All in all, we selected a total of 100 skills for our study.

\subsection{Alexa Skill Responses Crawler}

The most common interaction flow of an Alexa skill is the "open-command-stop" flow. 
To begin interacting with a skill, A user first says "Alexa, open X", where X is a skill's invocation name.
Then, the skill typically responds with an introduction or greeting message.
After that, the user starts uttering specific commands to make use of the skill's functionality. 
The skill responds with its answers or follow-up questions.
The conversation continues until the user says "Alexa, stop" to indicate their desire to quit. 
Sometimes, the skill responds with a goodbye message, but not always.
In order to study our research questions, we needed to have conversations like this with each of the 100 skills in our sample, recorded how each skill responded, and analyzed whether its responses followed or violated certain design guidelines.
Our initial attempt was fully manual.
Given a skill, we spoke to it, listened to and wrote down its responses in an excel spreadsheet, and coded the responses with respect to their compliance with design guidelines.
However, after about 20 skills, we found manual data collection time-consuming, difficult to scale to a large sample, and hard to replicate for other researchers.
Thus, we were motivated to develop a method to automate certain parts of this process.

We present a crawler tool we developed to automatically converse with a given skill and record the skill's responses. 
The input to this tool is a list of skill names.
The output is an excel spreadsheet containing each skill's responses (automatically recorded and transcribed) in various simulated conversation sessions.
Researchers can then review and analyze the spreadsheet data for their own research questions, which in our case are what design guidelines are more frequently adopted (RQ1) and how such adoption varies across categories (RQ2).



\begin{figure}[htb]
\includegraphics[width=3.4in]{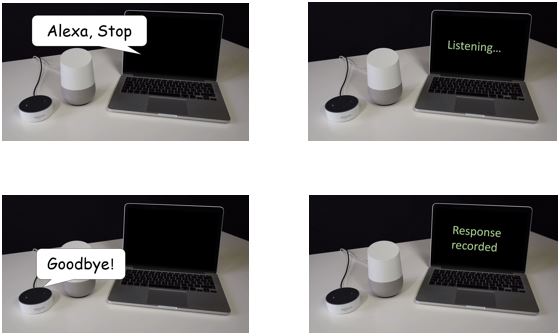}
\caption{Working Process of Our Voice Crawler}
~\label{fig:1}
\vspace{-0.6cm}
\end{figure}

Figure~ \ref{fig:1} provides a conceptual example of how our crawler simulates a voice conversation between users and Alexa devices.
To simulate speaking a command to a skill, our tool uses Google's Text-to-Speech package for Python\footnote{The project website is: https://gtts.readthedocs.io/en/latest/}. Then, it listens to and records the skill's responses. We used the Speech Recognition package for Python \footnote{The project website is: https://pypi.org/project/SpeechRecognition/} to implement the listening ability.
Given our sample of 100 skills, our crawler iterated through them, carried out a range of conversations with each skill, and listened to the skill's responses. 
This automatic data collection process is described pragmatically as follows:

\begin{algorithm}
\caption{Collect Responses to $m$ Commands by $n$ Skills}

\begin{algorithmic}[1]

     \FOR{\textit{skill} \textbf{in} [$s_1$,$s_2$,...,$s_n$]}
     
        \STATE speech $\gets$ TextToSpeech("Alexa, open \{\{skill's name\}\}");
    
        \STATE \textbf{play} speech
        
        \FOR{\textit{command} \textbf{in} [$c_1$,$c_2$,...,$c_m$]}
        
            \STATE speech $\gets$ TextToSpeech(\textit{command});
        
            \STATE \textbf{play} speech;
            
            \STATE audio $\gets$ \textbf{listen};
            
            \STATE text $\gets$ SpeechToText(audio);
            
            \STATE \textbf{save} text;
            
        \ENDFOR
        
    \ENDFOR
    
\end{algorithmic}

\end{algorithm}

\subsection{Guideline-Specific Response Elicitation Design}

Amazon's voice design guide \cite{GuideA} provides more than two dozens guidelines.
In our study, we limited the scope to a sample of eight guidelines.
They are denoted as G1 to G8 in the rest of the paper. 
For each design guideline, we needed to come up with an appropriate testing conversation flow in order to elicit responses we can evaluate, with respect to that guideline.
The details are presented below.

\textbf{Basic commands support (G1, G2, G3):} Three design guidelines recommend voice skills should support users to start, get help on, and end an interaction.
For Alexa skills, these translate into the ability to understand the basic commands of "open", "help", and "stop" respectively.
Described in more details, when a user says "Alexa, open [skill's name]," the skill needs to remain open and wait for the user's responses (G1); when a user asks "Alexa, help," the skill is expected to provide informative instructions such as introducing its core functionality (G2); and when a users says "stop," the skill should end the conversation naturally and gracefully with few or no words (G3).
In order to evaluate the compliance situation of these 100 skills with respect to G1, G2, G3, we designed crawler loops by setting the basic commands as elicitation commands. Within one round of the crawler loop, the crawler will say "open", "help", "stop" commands and listen to the responses in turn (this crawler loop is denoted as "open-help-stop" loop in the rest of the paper). Based on the responses collected, we would get to know how many skills support basic commands and conduct our analysis.
    

\textbf{Variety support (G4, G5):} Two design guidelines recommend voice skills should provide varying responses to the "open" (G4) and "stop" (G5) commands so that the interaction can feel more natural and less robotic.
To test whether a given skill complies with these guidelines, our 
crawler tool carried out an "open-help-stop" dialogue.
This dialogue was repeated $N$ times where $N$ is default to three so that we can detect variations in the skill's responses to "open" and "stop" command, if any. Particularly, in the first run of this dialogue, the tested skills was first-time enabled. The second round was run after solving all the account linking, age verifying steps. The third round was run after the skill has been fully explored (We manually conducted an "open-[commands]-stop" dialogue to fully explore it where commands are what the specific skill can support.
The list of commands were manually extracted from the skill's response to the "help" command in the second round.
For example, the skill \textit{"Examining the Scriptures Daily"} responded to "help" with the message \textit{"You can say tell me my daily text for today or read me my daily text for last Monday. You can also say read me tomorrow's daily text."}
We 
extracted three commands from this message: 
"tell me my daily text for today", "read me my daily text for last Monday" and "read me tomorrow's daily text."
Then each of the extracted commands was given to the skill.


\textbf{Error handling support (G6, G7):} 
We chose to evaluate two guidelines regarding error handling. 
The first guideline is when a skill receives no answer from a user regarding a question, it should deliver a re-prompt 
(G6). 
The second guideline is the re-prompt should be reworded or with more detailed instructional information(G7). 
To collect responses for a given skill regarding its error handling ability, our crawler first carried out a "open-help-stop" loop and repeated the "open-command-stop" loop three times to make sure the skill was fully explored (the commands were 
as how we handled G4, G5). After that , we enabled this skill again and stopped giving further command to wait for how it would respond.
Our crawler then repeated this process for all skills in our sample.

\noindent \textbf{Memorizing support (G8):} According to the design guide, users would appreciate it if a skill can remember their past interactions and provide more personalized services (G8).
In order to test this, 
our crawler first fully explored a skill's capabilities (like G4 and G5), and then carried out an "open-help-stop" loop one more time to see whether the skill remembered its last interaction and personalized its responses accordingly.


\subsection{Analyzing Responses by Design Guidelines}

Given our sample of 100 skills and 8 design guidelines to test for, our crawler automatically collected more than 1000 responses (the dataset is included in the supplementary material).
We manually analyzed the data as follows. 

\subsubsection{Data Correction}
First, we compared this dataset to a small pilot dataset of 20 skills we previously collected by hand in order to identify any discrepancy between machine and human transcribed responses. 
In doing so we were able to detect and correct problems brought by limitations of speech-to-text technology, such as typo and missing punctuation.

\subsubsection{Data Coding}
After data correction, two researchers independently coded each response's compliance with respect to design guidelines.
Afterwards, two researchers compared their coding results and resolved their discrepancies. 

For basic commands support, we examined collected responses to see whether the skill successfully executed the commands. 
For variety support, we compared the responses across repeated dialogues. If there are variations, we would code as following G4 or G5.
For error handling support, we first determined if the skill supports G6 and then compared with previous messages to determine whether the re-prompt messages were reworded or not.
Finally, for memorizing support, by comparing the last and the very first "open-help-stop" loop's responses, we judged whether the skill memorized previous interaction. 
The contents of the two responses were compared. If the second time's contents include any personalized information or previous interaction information while the first time doesn't, we would code as following G8.


\subsubsection{Comparative Analysis}

After we coded all the responses, we were finally able to address our research questions by comparing the results across guidelines (RQ1) and across categories (RQ2). 
For each guideline, we counted the number of skills that follow it and picked out both positive and negative examples for further investigation.
By comparing across different guidelines, we were able to understand the guideline adoption pattern in the wild, that is, which guidelines are obeyed by more or fewer skills.
By comparing across categoies, we were able to examine whether category can be a factor associated with whether certain guidelines are followed (or violated).

\section{Findings}
In this section, we present our findings regarding the current compliance situation of a sample of 100 skills with respect to eight voice design guidelines, in order to address the two research questions proposed earlier in the introduction.
In the following Improving Design Guidelines section, we will discuss several illuminating real-world examples, both positive and negative, we discovered during the data collection process, which serve to motivate further design recommendations for voice skills.

As described before, we initially selected 100 most popular skills from 10 different categories as our sample. All the responses were collected in late 2018.
However, after all the data was collected and cleaned, we needed to exclude six skills for which we failed to obtain meaningful results because of issues related to account linking or access permission.  
Hence, our findings presented below are based on 94 skills. 
\footnote{Our dataset can be accessed on request.}

\subsection{Basic Commands Support}


\subsubsection{Open Command (G1)}

According to the design guide and Amazon's Alexa building requirements \cite{invocation}, every skill we tested is expected to support open command (G1). When a user invokes a skill without specific intents (e.g. "Alexa, open [skill name]"), the skill is supposed to remain open and wait for the user's responses. At the same time, a welcome message which could also prompt the customer to continue interaction is also required.
We found all 94 skills supported G1.

\subsubsection{Help Command (G2)}
The "Help" command is used to help customers navigate a skill's core functionality. G2 states that every Alexa skill should implement the built-in "help" intent to provide better user experiences. 
We found only 81 out of 94 skills supported G2.
This left thirteen skills not supporting G2, including eight in the audio/music/sound category like \textit{4AFart} (a skill that plays fart sounds) , four one-shot \cite{oneshot} skills (skills that only involve single turn interactions) and one skill, \textit{Escape the Room}, in the Game category.

What could be the reasons these skills do not support G2?
One reason is that audio/music skills are meant for passive listening, as in the case of \textit{NPR One} and \textit{Thunderstorm Sounds}.
Another reason is that 
some skills only involve one-shot interactions
where a user asks a question or gives a command, the skill responds with an answer or confirmation, and the interaction is complete \cite{Alexa2018-lq}. 
Since one-shot skills will end the interaction and exit automatically after answering open utterance, users do not have a chance to say more commands, including the help command. 
Fact skills (skills that randomly tell users a fact concerning a certain topic when invoked) like \textit{Cat Facts} are good examples of these one-shot skills.
Furthermore, some other skills provide instructive information through other ways rather than a help message, as in the case of \textit{Escape the Room} from the Games category, which asks users to go to a website for reference in its opening message.

\subsubsection{Stop Command (G3)}

G3 states that every skill should respond to a user's "Alexa, stop" command. 
After the stop command is heard, a skill should exit and optionally return a response that is appropriate for the skill's functionality, such as a goodbye message \cite{Alexa2018-lq}. 
We found all 94 skills could successfully exit. 
Also, 74 of them gave a goodbye message.
For those skills who did not provide goodbye messages, most of them are one-shot skills which exit automatically after an one-sentence response. 


\subsection{Variety Support}

Compared to basic commands support, we found variety support is provided by much fewer number of voice skills in our sample. Below we present our findings for the two relevant design guidelines we studied.

\subsubsection{Variety in open responses (G4)} 

When a customer invokes a skill without specific intents ("Alexa, open [skill name]"), the skill should deliver an opening prompt. Skills are expected to provide several variations of opening prompts including one for first-time use, one for return and personalized prompts (G4).
We found 34 out of 94 skills (36\%) supported opening prompt variations. 
Furthermore, we observed they often served three use scenarios (with overlaps).
1. Some (n=8) were daily used skills or skills with regular updates; variety in opening prompts help keep users feel fresh and updated.
2. Some (n=16) were kills that remembered previous interactions; whenever users open the skill, its opening prompts will tell users where they left off last time.
3. Some (n=13) were skills with multiple states; the opening message will always inform users the current state. 

For the first scenario (daily use), one good example is the \textit{"Zyrtec"} skill which can report weather, pollen count and predominant allergens in a user-defined location.
When this skill was opened the first time, its opening prompt was \textit{"Hello! Let's get ahead of your allergies with today's Allergycast based on your location. Just follow these steps. One, Open your Alexa app on your phone ... [19 more words]"} 
For the second time, the skill said \textit{"let's start with your city and state, then we can get ahead of those allergies by setting up your allergy test report. What's your city and state?"}. 
For the third time, after the location was set, the skill's opening prompt turned into \textit{"Welcome to Zyrtec. Today in xxx, the pollen count is High, at 9.2 out of 12... [34 more words]"}
Comparing these three opening prompts, we found that when the skill was first enabled or used, it provided instructions about setting up step by step and elaborated clearly about the location requirement. 
After the skill got the location permission, the opening prompt changed into daily report of pollen. 
The whole interaction was natural and personal for users. 
In contrast, a poor example is \textit{Examining the Scriptures Daily}.
We found the skill always responded with the same sentence: \textit{``Which day do you like to hear''}. 
Although this opening prompt provided users with a cue to begin speaking and coached users on what to say next, the interaction could feel monotonous and less natural.

For the second scenario (remembering previous interactions), we found 26 of 94 skills (28\%) could remember previous interactions but only 16 supported variations (17\%).
A good example is \textit{7-min Workout} in the Health \& Fitness category. 
This skill is used to play instructions and background music for people who work out.
When the skill was firstly used, its opening message was \textit{``Welcome to Seven Minute Workout. When you are ready, just say start workout.''}.
After a previous workout was interrupted, the skill's opening message changed into \textit{``Welcome to Seven Minute Workout. To continue where you last left off, say ready. Otherwise, just say start workout.''} 
In this situation, variety in opening prompts provides a personalized experience for users by allowing them to pick up where they left off.
 
For the third scenarios (multiple states), 
a good example was the popular \textit{Magic Door} skill in the Games category.
This skill always informs users the current game state in the beginning so that users can choose to resume or restart the game. 
A negative example is \textit{Categories Game} skill, also in the Games category. 
The skill always says \textit{``howdy. You're playing Categories Game! For instructions, say help me or, say start playing!''} in its welcome message and have to start the game all over again no matter how many times it has been played.

\subsubsection{Variety in stop responses (G5)} 

According to \cite{Sciuto2018-ri}'s work, "Alexa, stop" is the most frequently used command. In this case, variety in stop responses could help users feel less like talking to a machine. 
We expected all the selected skills could add variety in their stop responses. However, based on our evaluation results,
We found most skills had really short goodbye messages like  \textit{"OK"} and \textit{"Goodbye."}. Only 19 of 96 skills (20\%) varied its goodbye messages. 
One good example is again the \textit{Zyrtec} skill. 
We found several variations such as 
\textit{"Ok. If you need allergy info, \textbf{I am here for you. Unless you move me.} Then I am over there for you. If you need to stock up on Zyrtec, just say my name, then order Zyrtec."}
, 
\textit{"Ok, if you need allergen information, \textbf{I will be here for you}. Remember, if you need to stock up on Zyrtec, I can help. Just say my name, then order Zyrtec"}
, and  
\textit{"Ok. When you need allergen information, \textbf{I am here 24 7 365}. If you need to stock up on Zyrtec, just say my name, then order Zyrtec."}
From these responses, we can see that although they expressed fairly the same meaning, the different wordings made the experiences felt less monotonous.

\subsection{Error Handling Support}

Error handling is an important part in any user interface design, voice skill design is no exception. 
In our study, we focused on a typical error handling scenario: when a customer responds to a skill prompt with silence. Under this situation, the skill is expected to deliver a re-prompt (G6) with rewording (G7) to disambiguate or elaborate on the kind of responses supported. Our findings are as follows.
\subsubsection{Re-prompting (G6) with Rewording (G7)}
For G6, We found a high percentage of skills supporting re-prompting---74 of 94 (79\%).

For G7, however, we found a low percentage of skills---23 of 94 (24\%)---that reword in their re-prompts.
%
A good example is the \textit{Amazon Story Time} skill. First, the skill greets users by saying \textit{"Welcome back to Amazon Story time! Would you like to resume The Mouse and the Unicorn?"}. After the question, if it receives no responses from the user, it would re-prompt with \textit{"you can say yes to resume or no to play the next story"}. 
In this example, the re-prompt offered more specific instruction for users to say yes or no. 
In contrast, a negative example is the \textit{Bring} skill in the Shopping \& Finance category, which simply stayed silent,
without any instruction or hint to help users handle a potential error.

\subsection{Memorizing Support (G8)}

Just like conversing with a friend, users appreciate when Alexa remembers what happened previously and what was said, especially for frequent actions and static information.
We found 27 of 94 skills that provided memorizing support. 

\noindent One positive example is \textit{Lemonade Stand} in the Kids category. 
It is a game where users can sell products and manage their income. 
The skill always remembers how the game ended  last time.
Each time a conversation began, this skill would say \textit{"Today is your twelfth day selling lemonade. Currently, it's windy and cool with some clouds. The forecast is a very low chance it will be warm and partly cloudy. Your cost for lemonade is fifteen cents a cup. You have \textbf{three dollars and fifty cents}. How many cups do you want to sell? "}
This message conveyed  the key statistics to help users remember their progress. 
In contrast, the \textit{5-min Plank Workout} skill in the Health \& Fitness category did not say anything explicitly to users that it remembered what exercises users might have done in the previous session. It always asked users to start over again, which could be frustrating.
As we examined further, we identified certain legitimate exceptions past interactions were not remembered. 
For example, a skill like \textit{This Day in History} is designed to be relevant for that day where past interactions do not matter. 

\begin{table*}[htb]
\caption{The Rate of Compliance for 8 Design Guidelines}
\vspace{0.4cm}
\includegraphics[width=7in]{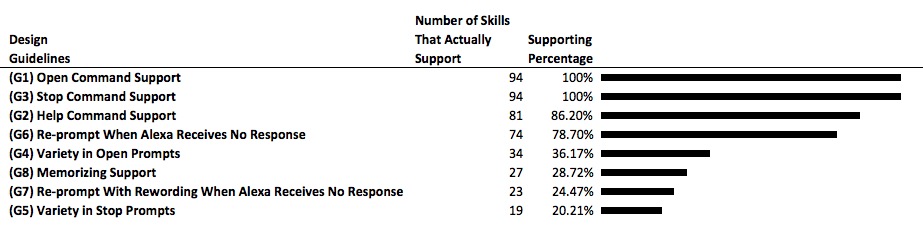}

~\label{tb:com_guide}
\end{table*}


\section{Comparative Analysis}

In previous sections, we presented our findings with respect to each of the eight guidelines 
(i.e., G1 to G8).
In this section, we will compare our findings across both guidelines and skill categories.
These comparisons address the two research questions in the introduction (i.e., RQ1 and RQ2).

\subsection{Across Design Guidelines (Q1)}
\label{cross-guidelines}
As shown in Table~ \ref{tb:com_guide}, we calculated the compliance rate for all 8 design guidelines and ranked them based on their rates. These results show that among the design guidelines we evaluated, some were more frequently violated than others.

Based on the ranking, open command support (G1) and stop command support (G3) were among those followed by the most number of skills. 
In contrast, Memorizing support (G8), rewording support (G7), and stop variation support(G5), were followed by the fewest skills.
As we can see, both G8 and G5 are related to Alexa skills' personalized services. 
What could explain such differences in compliance rate across design guidelines?
For guidelines related to personalized services, one possible explanation of their low adherence rate may be the difficulty in implementation, which involves user behavioral modeling, user data analysis and other techniques.
Also, high-quality personalized services require users to provide more personal information. 
It is hard to strike a good balance between the quality of personalized services and users' concern about their privacy \cite{Saffarizadeh2017-tm}. 
As for G7, the observed results told us that most of the skills only focused on providing re-prompts but did not take a further step to reword the repeated re-prompts to make a conversation more natural.

\subsection{Across Voice Skill Categories (Q2)}
\label{cross-categories}

In this part, we will make comparisons across different skill categories. 
As indicated in Table~ \ref{tb:com_cate}, we counted the number of design guidelines that a certain skill complied with and calculated the average number within one category.
Based on the calculated results, we obtained a ranking where the Games category was the top-ranked one while the Health \& Fitness and Entertainment categories had relatively low rankings.
Moreover, we broke down the comparison into four types of design guidelines: basic commands support (G1, G2, G3), variety support (G4, G5), error handling support (G6, G7) and memorizing support (G8). 
We first calculated the percentage of skills that supported each design guidelines within each category and then computed the average compliance rate within each of the four types. The results are also shown in Table~ \ref{tb:com_cate}.

\begin{table*}[htb]
\caption{The Average Support Rate in Four Design Features Across 10 Skill Categories}
\includegraphics[width=6in]{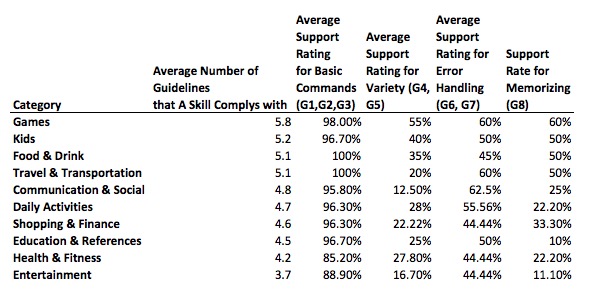}

~\label{tb:com_cate}
\end{table*}


From the ranking shown in Table~ \ref{tb:com_cate}, we can see that the Games category had the highest compliance rate for variety support guidelines (G4, G5), memorizing support guidelines (G8) and close to the highest compliance rate for error handling support (G6, G7).
Also can be seen is that the skills in the Game category were more likely to follow the rest of the guidelines. 
Game Skills are expected to involve more interactions with users and require more complicated user interface design, like remembering users' previous score and provide personalized game processes. 
When we looked at these 10 selected skills' user ratings on Amazon's website, they also achieved relatively high average user ratings (4.5/5), 
which matched with the comparative analysis results.
The skills in the Kids category also held high ranking positions in our table. One explanation is that children are considered a sensitive population that tends to have a higher requirement for design quality.

Let us now turn attention to categories with relatively low compliance rates.
Several interesting patterns emerged.
For example, the Entertainment category had close to the lowest compliance rate across all four guideline types.
At a quick glance, this finding was surprising because Entertainment and Games seemed similar yet occupied the two opposite ends in the ranking.
Upon closer examinations, we realized skills in the Entertainment category tend to offer quick and instant "fun" such as telling a joke or a compliment, 
 which do not need many user inputs and require less interaction design.
Another example is the Communication and Social category that had a relatively low compliance rate with respect to variety support guidelines but a high compliance rate for error handling guidelines.
One explanation could be that communication and social skills may involve users speaking longer and more intentional utterances and may be more prone to errors, which necessitates additional effort to handle errors.
In conclusion, with respect to RQ2, we found evidence that design guideline compliance patterns do differ greatly across categories, which suggests  associations between design guideline compliance and categories.
However, we were unable to determine whether these associations are causal or correlational, which will require further studies.

\newcommand{\user}[1]{\textbf{User:}  \textit{#1} \\}
\newcommand{\vgood}[1]{\textbf{Alexa:}  (Good) \textit{#1} \\}
\newcommand{\vbad}[1]{\textbf{Alexa:}  (Bad) \textit{#1} \\}
\newcommand{\vboth}[1]{\textbf{Alexa:}  \textit{#1} \\}


\section{Improving Design Guidelines}

Based on our findings and real skill examples we encountered during the evaluation process, we derived a set of new design guidelines for voice skills to complement the existing ones. In order to make our ideas clearer, simulated user-Alexa dialogues are presented for some of the points.

\subsection{Design Guidelines For One-Shot Skills}

We found that a significant number of instances of guideline violations are associated with one-shot skills such as \textit{Cat Facts} and \textit{Damn Girl}. 
In our sample, fewer than 10\% were one-shot skills but they accounted for a 
large number of guideline violations.
Upon closer examinations, some of the violations could be excused (users do not have chances to go deeper).
This observation may suggest that the official design guidelines need to be revised to consider the special needs of one-shot skills.
Here we present two ideas for the revision informed by our findings.

\subsubsection{Use Informative Invocation Name to Replace Help Command}
Since a one-shot skill often exits automatically after responding to users' commands, users may not have a chance to interact with the skill deeper. 
Thus, it is advisable to carefully choose an invocation name that is informative to remind users of its core functionality. 
A good example is the \textit{Rain Sounds} skill whose name clearly indicates that this skill intends to play rain sounds for users. 
In contrast, a negative example is the \textit{Damn Girl} skill, which carries a unusual name but gives users little information about what it does (in fact, it says a different compliment each time).

\subsubsection{Personalize the Contents Based on User's Interactions with Other Skills}
One-shot skills' interaction mode limits the collection of user inputs, which makes the process of personalizing their contents very difficult. 
In this case, a good way to solve the problem is to connect with other skills for more user inputs. 
For example, a one-shot skill aimed at providing basic facts about cats (such as \textit{Cat Facts}) could make use of a user's previous inquiries about a cat's health, collected from other skills. With this personalized information, this one-shot skill could provide more relevant health facts about cats the next time the user opens it.
But this approach must be implemented carefully to respect users' privacy preferences regarding sharing data across skills.

\subsection{Design Guidelines For Personalized Skill Services}
Our findings show that there is still a room for improvement in terms of providing personalized experiences for voice skill users.
During the process of analyzing our data, we noted several real-world design examples that could inform new design guidelines for voice skill developers.

\subsubsection{Change Interaction Mode for Repeat Users} 

Personalized service should not be limited to variety in responses, it should also be reflected through variations of the whole interaction mode. 
For example, through analyzing a user's interaction history, a skill could tell whether the user is a frequent user. 
If not, the skill could guide the user to explore its features in details. 
If yes, the skill could simplify or streamline the whole interaction flow to provide more personalized service.
For example, repeat users could get what they want immediately or receive a list of recommended services based on interaction history. 
Here we present an ideal interaction mode variation example. 
First is the interaction mode for non-frequent users. 

\noindent \user{Alexa, open Dishes Delivery.}
\vboth{OK, what kind of dishes do you want?}
...(the skill acquire necessary information like dishes kind, price, personalized taste like dishes cooked with no peppers)\\
\vboth{OK, got it. Your order is ready.}
\\
Next are the good and bad examples of the interaction mode for frequent users.\\
\user{Alexa, open Dishes Delivery.}
\vbad{OK, what kind of dishes do you want?}
\vgood{OK, welcome back. Do you still want "A" cooked with no peppers?}

\subsubsection{Providing detailed information via other platforms} 

One limitation of a voice skill is the amount of information it can provide in a single utterance.
Meanwhile, an overly long utterance in response to a user's question is highly discouraged.
In this case, we found some voice skills take advantage of other platforms such as mobile, emails, and SMS to deliver extra information.
A good example is the \textit{Store Card} skill. 
When this skill needs to tell users information that is not suitable through voice interaction, such as an URL, instead of saying it aloud, it sends the information to a user's mobile app and explains to the user that \textit{``we just made some improvements that you need to disable the skill and then enable it again. Please use the link we just sent to your app''}.
This practice eliminates the need for users to listen and remember long text.
Hence, we suggest that detailed information can be optionally provided via another platform.

\subsection{Other Design Guidelines}
Here we present several more design guidelines (not already covered by the official ones) informed by real world examples we observed, which reflected both good and bad design practices.


\subsubsection{Give feedback to help users locate problems in their commands}  
Users might feel frustrated when their commands cannot be correctly processed by a voice skill several times in a row.
Under this situation, if the skill could specifically tell users where the problems are in their input and give more specific instructions, it would more effectively help users adjust their input and receive the desired services from the skill. 
During the manual collection process, we found many skills just repeated the same generic sentence like \textit{"Sorry, I didn't understand that. What would you like?"} when researchers gave commands that could not be understood.
Those kind of responses do not provide any information about why Alexa cannot understand the user's command. 
We suggest an additional guideline that a skill should provide informative feedback such as telling users what it originally expected and why users' voice input did not match the expectation.
Here is an example dialogue contrasting a good response with a bad response with respect to this guideline.

\noindent \user{Alexa, open Pizza Delivery.}
\vboth{OK, what city do you live in?}
\user{My city is horse.}
\vbad{I didn't understand that. What city do you live in?}
\vgood{(The skill's logic does not think 'horse' is a city name.) Sorry, "horse" is not a city name, can you say your city's name again?}

\subsubsection{Let users know which skill they are currently interacting with}
Sometimes users may mistakenly think they are interacting with a skill but in fact with another skill.
They may say commands which are only meaningful for other skills but cannot be understood by this skill. 
In this case, a useful design guideline would be to remind users which skill they are interacting with when the skill fails to understand users.
The \textit{WebMD} skill is a good example following this guideline.
Below is a sample dialogue that demonstartes WebMD's informative response. 
\\
\\
(Suppose a user forgot to exit \textit{WebMD} but thought he is interacting with a pizza delivery skill.)\\
\user{Alexa, order pizzas.}
\vbad{Sorry, I didn't understand. What would you like to know?}
\vgood{Sorry, you are already speaking with the WebMD skill. You can ask things like "What is diabetes?" or "What are the side effects of Nexium?" What would you like to know?}

\subsubsection{Recognize and acknowledge problems in users' input} 

During our evaluation and the process of reviewing users' reviews, we noticed that for some skills, even if users give incorrect input, those skills still continue with the wrong information and respond to users with irrelevant answers. 
For example, \textit{Categories Game} is a skill that presents different categories and asks users to come up with a word that begins with a certain letter in each category. 
One of the reviews said that the skill sometimes does not seem to understand the words users actually spoke and continues the game regardless.
Hence, we suggest a skill should improve the ability to recognize different types of errors and acknowledge those errors, rather than acting if there is no error.

\subsubsection{Don't give advertisements or encouragements too often after "stop"} 

We observed that some skills include advertisements or encourage users to give ratings in the goodbye message too often. 
For example, the skill \textit{Big Sky} always asks users to write a review at termination.
One reviewer complained \textit{``A nice, helpful app, except that it frequently ends answers to queries with `please consider writing a review for this skill...', etc. I end up spending more time stuck listening to it beg for a review than, say, finding out what the temperature is.''} 
Whenever users give the stop command, they hope to stop the skill successfully rather than listen to other bunch of sentences. In this case, if the skill could reduce the frequency of or stop giving advertisements after user says ``stop'', it would provide better user experiences. 
Instead, skill developers should use other channels to do their advertising. 
    
\subsubsection{Don't include questions in goodbye messages}

We observed that some voice skills include a question in its goodbye message in response to the ``stop'' command.
For example, the \textit{Alexa Prize Socialbots} responds to a stop command by uttering \textit{``Thanks for chatting! Quick question. On a scale from 1 to 5 stars... how do you feel about speaking with this socialbot again?''}
We found this practice problematic.
As mentioned before, whenever users say ``stop'' to a skill, they indicate strongly that they do not wish to interact with the skill anymore.
But asking a question in the goodbye message would require users to continue the interaction, which may negatively affect the user experiences.
Hence, we suggest that voice skills hould not include questions in goodbye messages.

\section{HCI Research Agendas}


\subsection{Improving User Experiences}

Although current design guidelines and suggestions we offered before have already covered many design problems developers might meet, future research is still necessary to revise the guidelines to meet new needs.

First of all, research on understanding the design space of voice skills is critical. 
The design space of traditional voice user interfaces has been proposed \cite{Cohen2004-fm}, 
which includes three variables: 
    grammars---possible things users can say in response to each prompt and which are understood by the system), 
    dialog logic---actions taken by the system,
    and prompts---the recordings or synthesized speech played to the user during the dialog. 
However, the design space of the new generation of voice skills by third-party developers has not been adequately researched.
Although voice skill design and voice user interface design share certain problems such as how to express effective information through a natural and conversational interaction without graphical assistance, voice skills still have their own characteristics including strong interaction objectives, shared interaction features across ecosystems and so on. 
We argue that design guidelines for traditional voice user interfaces only serve as a good starting point for understanding the design space of voice skills.


Next, we found that connecting with other platforms is important for a voice skill, especially when the skill has versions on other platforms, such as \textit{Uber}, that have corresponding mobile versions. 
In order to improve user experiences, sharing information across different platforms is critical.
For example, an Alexa skill could give users a concise message while detailed information could be sent to users' mobile application.
Another benefit of connecting with other platforms is to allow users to receive consistent services. 
Hence, future studies are needed to understand how to best support a seamless cross-platform experience.

Finally, the design of personalized voice skills is also worth studying in the future. 
From our findings, we found personalized services do not have very high support rate among the current popular skills.
This finding implies personalized design requires more attention and further revision.
Studies show that in order to achieve a high degree of personalization, more personal or private information is often required from users. However, due to privacy concerns, users may want to disclose less personal information \cite{Saffarizadeh2017-tm, Cowan2017-zc}.
In this case, how to mitigate users' concerns that their privacy might be invaded can be an inspiring topic to be explored in future research. Methodologies related to investigating VUI users' privacy concern have been adopted in various research works \cite{Dubiel2018-bq, Cowan2017-zc}, which can be deployed in the future.

\subsection{Category-specific Design}

One contribution of this paper is the finding that there exists a high degree of variation in design guideline compliance across different skill categories. 
Thus, the variation we found suggests each skill category has its own specific requirements and design challenges.
This opens up several research questions for future, such as how should design guidelines and design space be adapted for different categories and even further, application scenarios? 
We can start from understanding which design guidelines are more important and needing more attention for each category. 
Also, we can study the variations in interaction flows across categories and understand the different challenges one may face in evaluating the skills in each category.

\subsection{Evaluation Methodology}


The evaluation methodology presented in this paper has several limitations.
On the technical side, our crawler is a research prototype that covers only selected design guidelines.
More research is still needed to support others.
Also, the speed of our tool is limited by the natural speed of a human's voice (since our tool simulates a human's interaction with a voice skill). 
In terms of data collection, we only focused on popular skills and categories. We do not yet know whether our findings can be generalized to less popular skills. 
At the same time, only 8 design recommendations were evaluated and we did not explore all the possible commands. Furthermore, the responses we collected were only one snapshot in time; we don't know whether tested skills have since updated their interaction models. 
In terms of the crawling algorithm, we cannot cover all the possible situations, like when we tested variety (G4,G5), the crawler only repeated the same commands for three times. 
The possibility of variety appeared in the fourth time or later was not eliminated. 
In terms of responses analysis, all the response labeling was conducted manually, which leaves room for improvement.

Correspondingly, there exist several possibilities for improving the evaluation methodologies in the future in order to better triangulate usability issues and design guideline violations. 
They include increasing the size and variety of the response data collection, integrating log data analysis (user's interaction history) \cite{Sciuto2018-ri} to enrich the commands and responses dataset, and automating the labeling and evaluation process. 

\section{Conclusion}
With the popularity of customized voice services, evaluation on them is of more importance than ever before. 
In our paper, we conducted design evaluation of a sample of 100 most popular Alexa skills from ten different categories using a voice skill crawler. 
The entire evaluation was performed with respect to eight design guidelines. 
Our findings revealed how these selected skills followed the guidelines. 
Based on our findings and the real sample responses we encountered during the evaluation process, we made several suggestions for improving the design of voice skills and identified challenges as well as opportunities for future research.

\balance{}

\bibliographystyle{SIGCHI-Reference-Format}
\bibliography{references}


\begin{thebibliography}{00}


\ifx \showCODEN    \undefined \def \showCODEN     #1{\unskip}     \fi
\ifx \showDOI      \undefined \def \showDOI       #1{{\tt DOI:}\penalty0{#1}\ }
  \fi
\ifx \showISBNx    \undefined \def \showISBNx     #1{\unskip}     \fi
\ifx \showISBNxiii \undefined \def \showISBNxiii  #1{\unskip}     \fi
\ifx \showISSN     \undefined \def \showISSN      #1{\unskip}     \fi
\ifx \showLCCN     \undefined \def \showLCCN      #1{\unskip}     \fi
\ifx \shownote     \undefined \def \shownote      #1{#1}          \fi
\ifx \showarticletitle \undefined \def \showarticletitle #1{#1}   \fi
\ifx \showURL      \undefined \def \showURL       #1{#1}          \fi

\bibitem{Jresearch}
 2018.
\newblock The Rise Of Chinese Voice Assistants And The Race To Commoditize
  Smart Speakers.
\newblock   (2018).
\newblock
\showURL{%
\url{https://www.cbinsights.com/research/china-voice-assistants-smart-speakers-ai/}}


\bibitem{invocation}
{Amazon Alexa}. 2018a.
\newblock Choose the Invocation Name for a Custom Skill.
\newblock   (2018).
\newblock
\showURL{%
\url{https://developer.amazon.com/docs/custom-skills/choose-the-invocation-name-for-a-custom-skill.html}}


\bibitem{AA18}
{Amazon Alexa}. 2018b.
\newblock Test and Submit Your Skill for Certification.
\newblock   (2018).
\newblock
\showURL{%
\url{https://developer.amazon.com/docs//devconsole/test-and-submit-your-skill.html}}


\bibitem{GuideA}
{Amazon Alexa}. 2018c.
\newblock Voice Design Guide.
\newblock   (2018).
\newblock
\showURL{%
\url{https://developer.amazon.com/designing-for-voice/}}


\bibitem{Alexa2018-rd}
{Amazon Alexa}. 2018d.
\newblock Voice Experiences | Alexa Design Guide.
\newblock
  \url{https://developer.amazon.com/docs/alexa-design/voice-experience.html#how-to-use-voice-input}.
    (July 2018).
\newblock
\newblock
\shownote{Accessed: 2019-1-18.}


\bibitem{oneshot}
{Amazon Alexa}. 2018e.
\newblock Voice Interface and User Experience Testing for a Custom Skill.
\newblock   (2018).
\newblock
\showURL{%
\url{https://developer.amazon.com/docs/custom-skills/voice-interface-and-user-experience-testing-for-a-custom-skill.html}}


\bibitem{Alexa2018-lq}
{Amazon Alexa}. 2018f.
\newblock Voice Interface and User Experience Testing for a Custom Skill |
  Custom Skills.
\newblock
  \url{https://developer.amazon.com/docs/custom-skills/voice-interface-and-user-experience-testing-for-a-custom-skill.html#46-one-shot-phrasing-for-sample-utterances}.
    (July 2018).
\newblock
\newblock
\shownote{Accessed: 2019-1-18.}


\bibitem{Ali-Hasan18}
{Noor Ali-Hasan}. 2018.
\newblock Evaluating Smartphone Voice Assistants: A Review of {UX} Methods and
  Challenges.
\newblock   (2018).
\newblock
\showURL{%
\url{https://voiceux.files.wordpress.com/2018/03/ali-hasan.pdf}}


\bibitem{Alexaskill}
{Corey Badcock}. 2015.
\newblock First Alexa Third-Party Skills Now Available for Amazon Echo.
\newblock   (2015).
\newblock
\showURL{%
\url{https://developer.amazon.com/blogs/post/TxC2VHKFEIZ9SG/First-Alexa-Third-Party-Skills-NowAvailable-for-Amazon-Echo}}


\bibitem{Braun2017-yo}
{Michael Braun}, {Nora Broy}, {Bastian Pfleging}, {and} {Florian Alt}. 2017.
\newblock \showarticletitle{A Design Space for Conversational In-vehicle
  Information Systems}. In {\em Proceedings of the 19th International
  Conference on {Human-Computer} Interaction with Mobile Devices and Services}
  {\em (MobileHCI '17)}. ACM, New York, NY, USA, 79:1--79:8.
\newblock


\bibitem{Brown2017-ao}
{Fred~A Brown}, {Mitchell~G Lawrence}, {and} {Victor~O'brien Morrison}. 2017.
\newblock Conversational virtual healthcare assistant.
\newblock   (Jan. 2017).
\newblock


\bibitem{Chen2015-lb}
{Hsuan-Eng Chen}, {Yi-Ying Lin}, {Chien-Hsing Chen}, {and} {I-Fang Wang}. 2015.
\newblock \showarticletitle{{BlindNavi}: A Navigation App for the Visually
  Impaired Smartphone User}. In {\em Proceedings of the 33rd Annual {ACM}
  Conference Extended Abstracts on Human Factors in Computing Systems} {\em
  (CHI EA '15)}. ACM, New York, NY, USA, 19--24.
\newblock


\bibitem{Cohen2004-fm}
{Michael~H Cohen}, {Michael~Harris Cohen}, {James~P Giangola}, {and} {Jennifer
  Balogh}. 2004.
\newblock {\em Voice User Interface Design}.
\newblock Addison-Wesley Professional.
\newblock


\bibitem{Cowan2017-zc}
{Benjamin~R Cowan}, {Nadia Pantidi}, {David Coyle}, {Kellie Morrissey}, {Peter
  Clarke}, {Sara Al-Shehri}, {David Earley}, {and} {Natasha Bandeira}. 2017.
\newblock \showarticletitle{``What Can {I} Help You with?'': Infrequent Users'
  Experiences of Intelligent Personal Assistants}. In {\em Proceedings of the
  19th International Conference on {Human-Computer} Interaction with Mobile
  Devices and Services} {\em (MobileHCI '17)}. ACM, New York, NY, USA,
  43:1--43:12.
\newblock


\bibitem{Damacharla2018-in}
{Praveen Damacharla}, {Parashar Dhakal}, {Sebastian Stumbo}, {Ahmad~Y Javaid},
  {Subhashini Ganapathy}, {David~A Malek}, {Douglas~C Hodge}, {and} {Vijay
  Devabhaktuni}. 2018.
\newblock \showarticletitle{Effects of {Voice-Based} Synthetic Assistant on
  Performance of Emergency Care Provider in Training}.
\newblock {\em International Journal of Artificial Intelligence in Education\/}
  (March 2018).
\newblock


\bibitem{Googleactions}
{Jason Douglas}. 2016.
\newblock Start building Actions on Google.
\newblock   (2016).
\newblock
\showURL{%
\url{https://developers.googleblog.com/2016/12/start-building-actions-on-google.html}}


\bibitem{Dubiel2018-bq}
{Mateusz Dubiel}, {Martin Halvey}, {and} {Leif Azzopardi}. 2018.
\newblock \showarticletitle{A Survey Investigating Usage of Virtual Personal
  Assistants}.
\newblock  (July 2018).
\newblock


\bibitem{Fytrakis2015-td}
{Emmanouil Fytrakis}, {Ioannis Georgoulas}, {Jose Part}, {and} {Yuting Zhu}.
  2015.
\newblock \showarticletitle{Speech-based Home Automation System}. In {\em
  Proceedings of the 2015 British {HCI} Conference} {\em (British HCI '15)}.
  ACM, New York, NY, USA, 271--272.
\newblock


\bibitem{Gheran2018-od}
{Bogdan-Florin Gheran}, {Jean Vanderdonckt}, {and} {Radu-Daniel Vatavu}. 2018.
\newblock \showarticletitle{Gestures for Smart Rings: Empirical Results,
  Insights, and Design Implications}. In {\em Proceedings of the 2018 Designing
  Interactive Systems Conference} {\em (DIS '18)}. ACM, New York, NY, USA,
  623--635.
\newblock


\bibitem{Voiceskillnumber}
{Bret Kinsella}. 2018.
\newblock Amazon Alexa Skill Count Surpasses 30,000 in the U.S.
\newblock   (2018).
\newblock
\showURL{%
\url{https://voicebot.ai/2018/03/22/amazon-alexa-skill-count-surpasses-30000-u-s/}}


\bibitem{Voice18}
{Bret Kinsella} {and} {Ava Mutchler}. 2018.
\newblock Smart Speaker Consumer Adoption Report 2018.
\newblock   (2018).
\newblock
\showURL{%
\url{https://voicebot.ai/wp-content/uploads/2018/03/smart_speaker_consumer_adoption_report_2018.pdf}}


\bibitem{Luger2016-pd}
{Ewa Luger} {and} {Abigail Sellen}. 2016.
\newblock \showarticletitle{Like having a really bad {PA}: the gulf between
  user expectation and experience of conversational agents}. In {\em
  Proceedings of the 2016 {CHI} Conference on Human Factors in Computing
  Systems}. dl.acm.org, 5286--5297.
\newblock


\bibitem{Luria2017-uz}
{Michal Luria}, {Guy Hoffman}, {and} {Oren Zuckerman}. 2017.
\newblock \showarticletitle{Comparing Social Robot, Screen and Voice Interfaces
  for {Smart-Home} Control}. In {\em Proceedings of the 2017 {CHI} Conference
  on Human Factors in Computing Systems} {\em (CHI '17)}. ACM, New York, NY,
  USA, 580--628.
\newblock


\bibitem{McReynolds2017-zx}
{Emily McReynolds}, {Sarah Hubbard}, {Timothy Lau}, {Aditya Saraf}, {Maya
  Cakmak}, {and} {Franziska Roesner}. 2017.
\newblock \showarticletitle{Toys That Listen: A Study of Parents, Children, and
  {Internet-Connected} Toys}. In {\em Proceedings of the 2017 {CHI} Conference
  on Human Factors in Computing Systems} {\em (CHI '17)}. ACM, New York, NY,
  USA, 5197--5207.
\newblock


\bibitem{Miniukovich2017-kd}
{Aliaksei Miniukovich}, {Antonella De~Angeli}, {Simone Sulpizio}, {and} {Paola
  Venuti}. 2017.
\newblock \showarticletitle{Design Guidelines for Web Readability}. In {\em
  Proceedings of the 2017 Conference on Designing Interactive Systems} {\em
  (DIS '17)}. ACM, New York, NY, USA, 285--296.
\newblock


\bibitem{GuideG}
{Actions on Google}. 2018.
\newblock Conversation Design.
\newblock   (2018).
\newblock
\showURL{%
\url{https://designguidelines.withgoogle.com/conversation/}}


\bibitem{Purington2017-ky}
{Amanda Purington}, {Jessie~G Taft}, {Shruti Sannon}, {Natalya~N Bazarova},
  {and} {Samuel~Hardman Taylor}. 2017.
\newblock \showarticletitle{Alexa is my new {BFF}: social roles, user
  satisfaction, and personification of the amazon echo}. In {\em Proceedings of
  the 2017 {CHI} Conference Extended Abstracts on Human Factors in Computing
  Systems}. dl.acm.org, 2853--2859.
\newblock


\bibitem{Saffarizadeh2017-tm}
{Kambiz Saffarizadeh}, {Maheshwar Boodraj}, {and} {Tawfiq~M Alashoor}. 2017.
\newblock \showarticletitle{Conversational Assistants: Investigating Privacy
  Concerns, Trust, and {Self-Disclosure}}. In {\em {ICIS} 2017 Proceedings}.
  aisel.aisnet.org.
\newblock


\bibitem{Sciuto2018-ri}
{A Sciuto}, {A Saini}, {J Forlizzi}, {and} {J~I Hong}. 2018.
\newblock \showarticletitle{Hey Alexa, What's Up?: A {Mixed-Methods} Studies of
  {In-Home} Conversational Agent Usage}.
\newblock {\em Proceedings of the 2018 on\/} (2018).
\newblock


\bibitem{Smith2016-zu}
{Taliesin~L Smith}, {Clayton Lewis}, {and} {Emily~B Moore}. 2016.
\newblock \showarticletitle{Demonstration: Screen Reader Support for a Complex
  Interactive Science Simulation}. In {\em Proceedings of the 18th
  International {ACM} {SIGACCESS} Conference on Computers and Accessibility}
  {\em (ASSETS '16)}. ACM, New York, NY, USA, 319--320.
\newblock


\bibitem{Vtyurina18}
{Alexandra Vtyurina}. 2018.
\newblock 5 Seconds After: Exploring User Actions with Voice Assistants in the
  Moments After a System Response.
\newblock   (2018).
\newblock
\showURL{%
\url{https://voiceux.files.wordpress.com/2018/03/vtyurina.pdf}}


\bibitem{Zou2015-jn}
{Hong Zou} {and} {Jutta Treviranus}. 2015.
\newblock \showarticletitle{{ChartMaster}: A Tool for Interacting with Stock
  Market Charts Using a Screen Reader}. In {\em Proceedings of the 17th
  International {ACM} {SIGACCESS} Conference on Computers \& Accessibility}
  {\em (ASSETS '15)}. ACM, New York, NY, USA, 107--116.
\newblock


\end{thebibliography}

\end{document}